\documentclass[journal]{IEEEtran}

\ifCLASSINFOpdf
\else
\fi
\usepackage{cite}
\usepackage{graphicx}
\usepackage{textcomp}
\usepackage{hyperref}
\usepackage{multirow}
\usepackage{adjustbox}
\usepackage[table,xcdraw]{xcolor}
\usepackage{xcolor}
\usepackage{amsmath}

\begin{document}
%
\title{mm-Pose: Real-Time Human Skeletal Posture Estimation using mmWave Radars and CNNs}
%
%
%
\author{\IEEEauthorblockN{\large{Arindam Sengupta, Feng Jin, Renyuan Zhang and Siyang Cao}}\\
\IEEEauthorblockA{{Department of Electrical and Computer Engineering, University of Arizona, Tucson, AZ USA} \\
Email: \{sengupta, fengjin, ryzhang, caos\}@email.arizona.edu}
}

\maketitle

\begin{abstract}
In this paper, \textit{mm-Pose}, a novel approach to detect and track human skeletons in real-time using an mmWave radar, is proposed. To the best of the authors' knowledge, this is the first method to detect $>$15 distinct skeletal joints using mmWave radar reflection signals. The proposed method would find several applications in traffic monitoring systems, autonomous vehicles, patient monitoring systems and defense forces to detect and track human skeleton for effective and preventive decision making in real-time. The use of radar makes the system operationally robust to scene lighting and adverse weather conditions. The reflected radar point cloud in range, azimuth and elevation are first resolved and projected in Range-Azimuth and Range-Elevation planes. A novel low-size high-resolution radar-to-image representation is also presented, that overcomes the sparsity in traditional point cloud data and offers significant reduction in the subsequent machine learning architecture. The RGB channels were assigned with the normalized values of range, elevation/azimuth and the power level of the reflection signals for each of the points. A forked CNN architecture was used to predict the real-world position of the skeletal joints in 3-D space, using the radar-to-image representation. The proposed method was tested for a single human scenario for four primary motions, (i) Walking, (ii) Swinging left arm, (iii) Swinging right arm, and (iv) Swinging both arms to validate accurate predictions for motion in range, azimuth and elevation. The detailed methodology, implementation, challenges, and validation results are presented.   
\end{abstract}


%
\IEEEpeerreviewmaketitle
\section{Introduction}
\IEEEPARstart{W}{ith} the advent in computing resources and advanced machine learning (ML) techniques, computer vision (CV) has emerged as an exciting field of research to provide Artifical Intelligence (AI) and autonomous machines with information about the visual representation of the real world \cite{forsyth2002computer,szeliski2010computer}. Primarily using vision based sensors, such as monocular camera, RGBD camera or IR based sensors, and applied machine learning, CV targets several applications, including (but not limited to) object classification, target tracking, traffic monitoring and autonomous vehicles\cite{krizhevsky2012imagenet,messelodi2005computer,petrovskaya2009model,dobrokhodov2006vision,reulke2007traffic}. In the recent years, another interesting topic that the CV community has been exploring is the ability to estimate human skeletal pose by identifying and detecting specific joints and/or body parts from still/video data. This specific area of research finds several applications, one being primarily in the health-care industry by automating patient monitoring systems, with the current situation of global shortage in nursing staff \cite{oulton2006global}. Such tracking systems would also allow for effective pedestrian monitoring for autonomous and semi-autonomous vehicles, and aid defense forces with behavioral information of the adversary, to trigger appropriate preventive decision making.  

\begin{figure}[t!]
\centering
\includegraphics{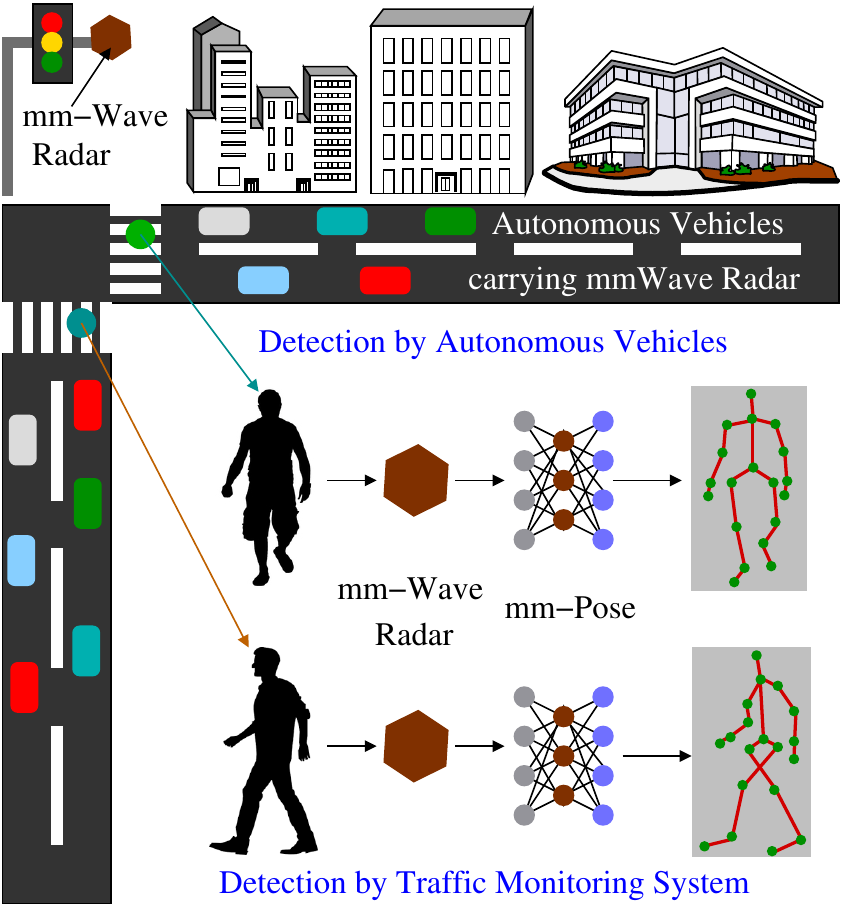}
\caption{\textit{mm-Pose} can be used in autonomous/ semi-autonomous vehicles and traffic monitoring systems for robust skeletal posture estimation of pedestrian.}
\vspace{-0.7cm}
\label{appl}
\end{figure}

While vision based sensors provide a high-resolution representation of the scene, there are a few challenges associated with their operation. They heavily rely on external sources for illuminating the scene and are therefore rendered ineffective in poor lighting conditions, adverse weather conditions or when the scene/target is occluded. These could result in irrevocable catastrophic events similar to the one encountered at the Uber self-driving vehicle crash incident in Arizona due to the vision/LiDAR sensors' inability to detect the pedestrian in time to avoid the accident. There is therefore an imminent need for alternate sensors to achieve the task, while overcoming the aforementioned challenges.

Radio Frequency (RF) based sensors, such as radars, use its own signals to illuminate the target, therefore making it operationally robust to scene lighting and weather conditions. However, unlike vision based sensors, radars only represent the scene with reflection point clouds rather than a true-color image representation. Radars are therefore primarily used for target localization applications. Furthermore, object classification becomes non-trivial with the point cloud data alone, and the lack of available labeled radar data-sets for this task makes it even more challenging. 

Traditionally, radar systems have been size and cost intensive primarily targeted to commercial and defense applications. However, continuing advancement in micro-electronics fabrication and manufacturing techniques, including Radio Frequency Integrated Circuits (RFICs), have significantly reduced the size and cost of electronic sensors making them more accessible to public \cite{robertson2001rfic, hasch2012millimeter, alhalabi2011design}. mmWave automotive radars are an example of this technology. They are low-power, compact and are extremely practical to deploy. Furthermore, mmWave radars provides us with a high resolution point cloud representation of the scene and have therefore emerged as one of the primary sensors in autonomous robots on a smaller scale, to more commercial applications such as autonomous vehicles. Higher operating bandwidths also allow mmWave radars to roughly generate the contour of human body without extracting facial information, thus preserving user privacy.

In this paper, we propose \textit{mm-Pose}, a novel real-time approach to estimate and track human skeleton using mmWave radars and convolutional neural networks (CNNs). A potential depiction of its application in traffic monitoring systems and autonomous vehicles is shown in Fig.~\ref{appl}. To the best of the authors' knowledge, this is the first method that uses mmWave radar reflection signals to estimate the real-world position of $>$15 distinct joints of a human body. \textit{mm-Pose} could also find applications in (i) privacy-protected automated patient monitoring systems, and (ii) aiding defense forces in a hostage situation. Radars carrying this technology on unmanned aerial vehicles (UAVs) could scan the building and map the live skeletal postures of the hostage and the adversary, through the walls, which wouldn't have been possible otherwise with vision sensors.

The paper is organized as follows. Section II summarizes the current skeleton tracking work carried out in the CV community and its extension to RF sensors. A concise background theory around the two fundamental blocks of the system, namely (i) radar signal processing chain and (ii) machine learning and neural networks is presented in Section III. The detailed approach, novel data representation and system architecture are presented in Section IV, followed by the experimental results and discussion in Section V. Finally, the study is summarized and concluded in Section VI.


\section{Literature Review}
\par It is extremely critical to accurately estimate and track human posture in several applications, as the estimated pose is key to infer their specific behavior. Since the last decade, scientists have been exploring various approaches to estimating human pose. One of the early works in 2005 was \textit{Strike a Pose}, proposed by researchers at Oxford, that would detect humans in a specific pose by identifying 10 distinct body parts/limbs using rectangular templates from RGB images/videos \cite{ramanan2005strike}.  A $k$-poselet based keypoint detection scheme was proposed in 2016, that uses predicted torso keypoint activations to detect multiple persons using agglomerative clustering \cite{gkioxari2014using}. Another approach was to use region-based CNN (R-CNN) to learn $N$ masks, to detect each of the $N$ distinct key-points to construct the skeleton from images, using a ResNet variant architecture \cite{he2017mask}. In 2016, \textit{DeeperCut}, an improved multi-person pose estimation model from \textit{DeepCut} was proposed that used a bottom up approach using a fine-tuned ResNet architecture that doubled the then estimation accuracy with a 3 orders of magnitude reduction in run-time \cite{insafutdinov2016deepercut, pishchulin2016deepcut}. A top-down approach to pose estimation was proposed by Google, that first identified regions in the image containing people using R-CNN, and then used a fully convolutional ResNet architectiture and aggregation to obtain the keypoint predictions, yielding a 0.649 precision on the COCO test-dev set \cite{papandreou2017towards}. Another extremely popular bottom-up approach for human pose estimation is \textit{OpenPose}, proposed by researchers at Carneigie Mellon University in 2017 \cite{cao2017realtime}. \textit{OpenPose} used Part Affinity Fields (PAF), a non-parametric representation of different body parts, and then associate them to individuals in the scene. This real-time algorithm had great results on the MPII dataset and also won the 2016 COCO keypoints challenge \cite{lin2014microsoft}. Also, the cross-platform versatility and open-source data-sets has led to \textit{OpenPose} being used as the most popular benchmark for generating highly accurate ground truth data-sets for training. 

While the aforementioned approaches paved the way towards human pose and skeleton tracking, they were limited to 2-D estimation on account of the images/videos being collected from monocular cameras. While monocular cameras provide high resolution information of the azimuth and elevation of the objects, extracting depth using monocular vision sensors is extremely challenging and non-trivial. To model a 3-D representation of the skeletal joints, \textit{HumanEva} dataset was created by researchers at the University of Toronto \cite{sigal2010humaneva}. The dataset was created by using 7 synchronous video cameras (3 RGB + 4 grayscale) in a circular array, to capture the entire scene in its field-of-view. The human subject was made to perform 5 different motions, and reflective markers were placed on specific joint locations to track the motion and a ViconPeak commercial motion capture system was used to obtain the 3-D ground truth pose of the body. Another approach to extract 3-D skeletal joint information is by using Microsoft Kinect \cite{zhang2012microsoft}. The Kinect consists of an RGB and infra-red (IR) camera that allows it to capture the scene in 3-D space. It used a per-pixel classification approach to first identify the human body parts, followed by joint estimation by finding the global centroid of the probability mass for each identified part. However the downsides of vision based sensors for skeletal tracking are the fact that their performance is extensively hindered in poor lighting and occlusion. Moreover, as previously introduced, privacy concerns restrict the use of vision based for several applications.

\par Studies have previously made use of micro-doppler radar signatures to determine human behavior using RF signals, however it did not provide spatial information of the subjects' locations \cite{ref_patientbehaviour,leobehavior}. Skeleton tracking using RF signals is a new and emerging area of research. RF based devices can be further classified into two categories, wearable and non-wearable. Wearable wireless sensors use Wi-Fi signals to track the location and velocity of the device, which indirectly represents the human. However, Wi-Fi signals cannot distinguish between different body parts and are therefore not suited for the proposed task of pose estimation. Non-wearable RF sensors, such a radar, can be traditionally used to localize the target in range and angle. \textit{RF-Capture}, proposed by researchers at MIT in 2015, was the first approach to identify several human body parts, in a coarse fashion, using FMCW signals and an antenna array, and then stitching the identified parts to reconstruct a human figure \cite{Adib:2015:CHF:2816795.2818072}. However, the design couldn't perform a full skeletal tracking over time. This was soon followed by \textit{RF-Pose}, proposed by the same research group in 2018, that used RF heat-maps obtained using two antenna arrays, one vertical and the other horizontal \cite{Zhao_2018_CVPR}. A teacher-student encoder-decoder architecture was used to estimate various granular key-points, which were then used to construct the skeletal pose. Finally, \textit{RF-Based 3D Skeletons}, used an FMCW signal with a 1.8 GHz bandwidth and a ResNet architecture to estimate the 3-D positions of keypoints followed by triangulation to estimate a 3-D model of a human skeleton. A circular array of vision sensors was used to capture the scene and Open-Pose skeletal data served as the ground truth for supervised training\cite{RFpose}.
\par In this paper, we propose \textit{mm-Pose}, a novel approach to use 77 GHz mmWave radars for human skeletal tracking. mmWave radars offer a greater bandwidth ($\approx$3 GHz), that in turn provides a more precise resolution. Furthermore, operating at 77 GHz allows it to capture even small abnomalities from the reflection surface, thus adding more granularity in terms of identifying more keypoints. Unlike the aforementioned approaches, mmWave radars are low-power, low-cost and compact, making it extremely practical for deployment. We make use of a forked-CNN architecture to predict $>$15 key-points and construct the skeleton in real-time. To obtain ground truth data, we parallely collect the keypoint locations using Microsoft Kinect on MATLAB API.

\section{Background Theory}
\subsection{Radar Signal Processing}
\par The mmWave radar transmits a frequency modulated continues wave (FMCW) chirp signal, and utilizes stretch processing \cite{ref_radarbook_richards} to get the beat frequency, which corresponds to the target's range. The Doppler processing across multiple chirps during one coherent processing interval (CPI) determines the Doppler frequency, which is related to the target's velocity. Mathematically, the \(n\)-th chirp during one CPI in complex form is given by: 
\begin{equation}
\begin{aligned}
x_n(t)=e^{j2\pi[f_0t + \frac{BW}{2T}t^2]},&\ nT\leq t <(n+1)T,\\
&\ \forall{n}\in[0,\ 1,\ ...,\  N-1].
\end{aligned}
\label{equ_TXChirp}
\end{equation}
where \(f_0\) is the chirp starting frequency, \(BW\) is the sweeping bandwidth, \(T\) is the duration of one chirp and \(N\) is the number of the chirps during one CPI. \(\frac{BW}{T}\)
is referred to as the chirp rate. The echo from a target is a time delayed version of the transmitting chirp. After stretch processing, the resulting baseband signal is given as: 
\begin{equation}
A_r\times e^{j2\pi{f_0}\tau_n}\times e^{j2\pi\frac{BW}{2T}(2\tau_nt-\tau_n^2)}.
\label{equ_Mixing}
\end{equation}
where \(A_r\) is the normalized received signal amplitude, which represents its reflectivity, and the \(\tau_n\) is the two round time delay between the radar and the target during the \(n\)-th chirp period, 
\begin{equation}
\tau_n=\frac{2(R_0-vnT)}{c}.
\label{equ_Delay}
\end{equation}
in which \(R_o\) is the initial distance, \(v\) is the target's radial velocity. The radar cross section (RCS), which represent the reflectivity of the target, can be solved by: 
\begin{equation}
\sigma=20\log_{10}(4\pi R_o^2 A_r).
\label{equ_rcs}
\end{equation}
\par The equation above shows the relationship between the normalized signal amplitude, and the corresponding power, being directly proportional to the radar cross section or the size of the target. As \(\tau_n\) is constant for one chirp in the range dimension, the baseband signal is a single frequency tone with respect to \(t\), also called the beat frequency, given by:
\begin{equation}
f_{Beat}=\frac{BW}{T}\tau_n\approx\frac{BW}{T}\tau=\frac{BW}{T}\frac{2R_0}{c}.
\label{equ_BeatFreq}
\end{equation}
\par The beat frequency resolution, that depends on the sampling time in one chirp, is expressed as:
\begin{equation}
\Delta f_{Beat}\geq\frac{1}{T}.
\label{equ_BeatFreqRes}
\end{equation}
\par From (\ref{equ_BeatFreq}) and(\ref{equ_BeatFreqRes}), the range resolution can be calculated as: 
\begin{equation}
\Delta R\geq\frac{c}{2BW}.
\label{equ_RangeRes}
\end{equation}

\par In the Doppler dimension, the data is sampled in the same position during each chirp across all the \(N\) chirps. Tis time, as \(t\) is constant, the baseband signal in (\ref{equ_Mixing}) is a single tone with respect to \(\tau_n\) after ignoring the smaller multiplicative term, and is expressed as:
\begin{equation}
e^{j2\pi f_0\tau_n}=e^{j2\pi f_0\frac{2(R_0-vnT)}{c}}.
\end{equation}
\par The obtained Doppler frequency is given by:
\begin{equation}
f_{Doppler}=-\frac{f_02v}{c}=-\frac{2v}{\lambda}.
\label{equ_DopplerFreq}
\end{equation}
where \(\lambda\) is the wavelength. The Doppler frequency resolution depends on the sampling interval in one CPI, and is represented as:
\begin{equation}
\Delta f_{Doppler}\geq\frac{1}{CPI}
\label{equ_DopplerFreqRes}
\end{equation}
\par Using (\ref{equ_DopplerFreq}) and (\ref{equ_DopplerFreqRes}), the obtained velocity resolution is expressed as: 
\begin{equation}
\Delta v\geq\frac{\lambda}{2*CPI}.
\label{equ_VelocityRes}
\end{equation}

\par
\begin{figure}[h!]
	\centering
	\includegraphics[width=\columnwidth]{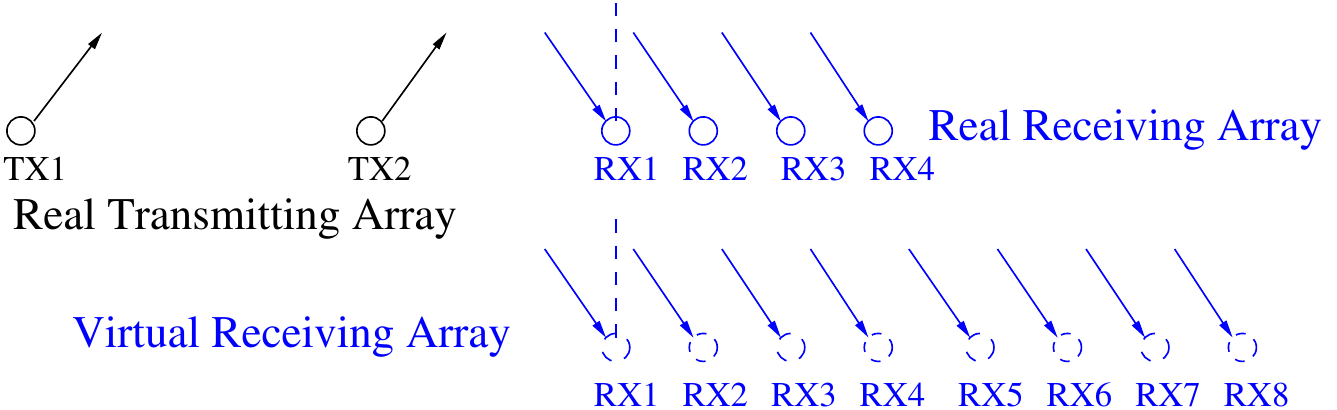}
	\caption{TDM-MIMO DOA.}
	\vspace{-0.1cm}
	\label{beamforming}
\end{figure}

 To determine the angle of the target, the time-division-multiplexing (TDM) multiple-input and multiple-output (MIMO) direction-of-arrival (DOA) estimation algorithm is used. Consider a scenario where a mmWave radar has two real transmitting antenna elements, and four real receiving antenna elements, as shown in Fig. \ref{beamforming}. First, TX1 transmits a chirp, and all the four real receiving antenna elements (RX1-RX4) recieve the echo with a progressive phase term \(\phi (n)\), depending on the angle-of-arrive (AOA) of the target \(\theta\). For the $n^{th}$ reciever element RX\(n\), we have:
\begin{equation}
\phi(n) = \frac{2\pi*n*dsin(\theta)}{\lambda}.
\label{equ_AOA}
\end{equation}
where \(d\) is the distance between two consecutive receiving elements (\(0.5\lambda\) in our case) to avoid the grating lobe effect.

After TX1 has transmitted a chirp, TX2 starts to transmit a chirp as well. Similar to the prior scenario, the receiving antenna elements recieve the echo with a progressive phase term. But as the distance between the TX1 and TX2 is \(4\lambda\), the phase terms get adjusted appropriately. If we view the signals from TX2 as being received by a ``virtual" receiving array indexing from 5 to 8, then the phase term for RX5 to RX8 will also be \ref{equ_AOA}. Once we obtain the 8 reciever samples with a progressive phase term, a simple FFT solves for the angle \(\theta\). The TDM-MIMO used here extends the size of receiving array from real size of \(3\lambda\) to a virtual size of \(7\lambda\), yielding a better angle resolution which is inversely proportional to the array size. Once we obtain the range ($R$) and AOA ($\theta$) of the desired target from the signal processing chain, we can then compute the cartesian distances - depth ($R\cos\theta$) and azimuth ($R\sin\theta$) of the target, with respect to the radar at origin.

\subsection{Neural Networks}
With the advent of graphical processing units (GPUs) and high-performance-computing (HPC), neural networks have emerged as one the most popular machine learning methods for classification and regression problems. Neural networks are loosely derived from biological neurons, where multiple neurons or nodes form an interconnected network to carry/transform the input signals to arrive at the desired output. Every node $i$ in a neural network accepts a weighted input and provides a non-linear output $O_i$ subject to a non-linear activation function, given by:
\begin{equation}
O_i = g_i(W_ix_i+b_i).
\end{equation}
where $W_i$ is the weight that the input $x_i$ is scaled by, $b_i$ is the bias and $g_i$ is the non-linear activation function. Without a non-linear activation function, a neural network would only result in a linear function estimation and would make it unsuitable for estimating complex non-linear relationships between the input and outputs. A neural network consists of three major sub-stages - (i) input layer, (ii) hidden layer(s) and (iii) output layer. The number of nodes in the hidden layers are equal to the number of features that we want the network to learn from, and the number of nodes in the output layers are equal to the number of classes (for classification problems) or number of desired outputs (for regression problems). The number of hidden layers and the number of nodes in a hidden layer are hyper-parameters that do not have a closed form expression, neither do we have prior information of what the outputs from each of the hidden layers should be. The objective of the hidden layers is to transform the input data to a higher-dimensional space in order to achieve the desired classification or regression task in hand. The appropriate transformation is established by learning the optimum values of the weights, that minimizes the desired loss function by a gradient descent algorithm using back-propagation. 

While traditional multi-layer-perceptrons (MLPs), as described above, are ideal for most tasks, convolutional neural networks (CNNs) empirically perform better for tasks involving images. Analogous to hidden layers, each CNN layer can represent multiple higher dimensional representation of the input images, based on the specified depth of the layer. For instance a CNN layer with depth 32 would generate 32 unique transformed representation of the input. The transformation is carried out via m$\times$m weight kernel. Similar to traditional nodes, the kernel would first take the weighted inputs (pixels), sum them, and then apply a non-linear activation function to yeild a single valued scalar as an output. This process is repeated when the kernel mask traverses the entirety of the image with a user-defined stride length. For instance, if an ${N\times N\times 3}$ image is subjected to a CNN layer with depth $D$, and a kernel size ${k\times k\times 3}$ ($k<N$), we would obtain a ${N\times N\times D}$ volume tensor as output, with $D$ distinct ${k\times k\times 3}$ weights to be trained. The training process is similar to MLPs, where gradient descent (or its optimized variants) using back-propagation is used.

Unlike the hidden layer outputs from MLP, the CNN ``filters" could have a visual representation. In some cases with a given visual input, say of a cat, the resulting filters detected the outline, edges, ears, nose etc. in the activation map, due to the fact than CNNs are inherently spatial filters. Another added advantage that CNNs offer over MLP is the significant reduction in computational complexity, owing to the kernel weights getting reused for generating a single transformation. As an example, if an ${N\times N\times 3}$ image is subjected to a CNN layer with depth $1$ (for simplicity), and a kernel size ${k\times k\times 3}$, the total number of trainable parameters would be $3k^2$, as opposed to $3N^2$ with MLP. With the increase in the number of representations $D$, while the additional number of parameters in CNNs would increase linearly ($D\times(3k^2)$), the increase in the number of parameters in a fully-connected MLP would increase exponentially ($(3N^2)^D$).    

\section{Proposed Approach}
\subsection{Radar-To-Image Data Representation}
As introduced in the previous sections, radars are essentially time-of-flight sensors that illuminate the scene with its own RF signals and use the phase information of the reflected signals to resolve the time-delay and estimate the range of the points of reflection. As the name suggests, mmWave radar signal wavelengths are in the order of mm, which enables them to even detect minute abnormalities of a target. Furthermore, with bandwidths in the range of 3-4 GHz, mmWave radars can also provide a high resolution mapping of the scene in range.

\begin{figure}[t!]
\centering
\includegraphics{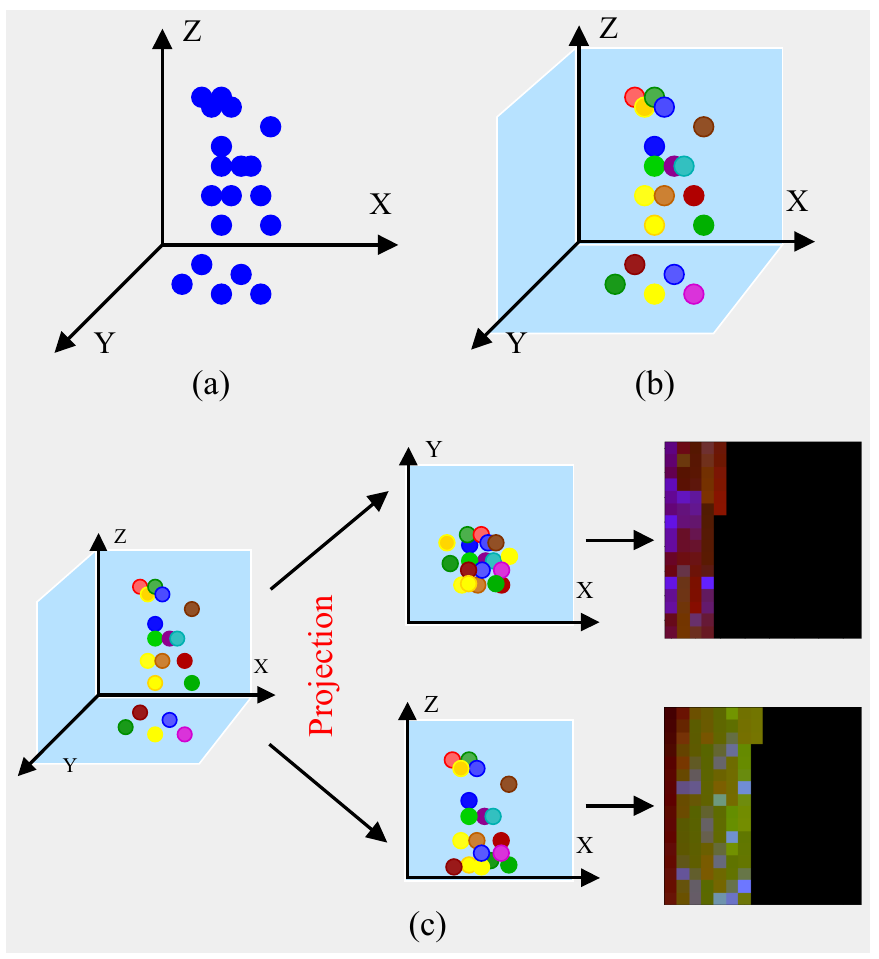}
\caption{(a) Point-cloud representation of the reflection points in 3-D space; (b) Point-cloud representation with the reflected power in the color channels; (c) Projecting the point-cloud with intensities in the XY and XZ planes, followed by the image equivalent with the RGB channels corresponding to X,Y/Z and Intensity respectively.}
\vspace{-0.45cm}
\label{pcl}
\end{figure}

The radar reflections over a coherent processing interval (CPI) results in a radar data cube with 3 dimensions, viz. fast-time, slow-time and channel. By using the radar signal processing chain, as described in Section~III-A, we obtain the range, velocity and angle information of the reflection points. By using basic trigonometric relations, we obtain the real world position $(x,y,z)$ of the reflection points, with respect to the radar (at origin), where ${x,y,z}$ represent the depth, azimuth and elevation coordinates, respectively. In this study, we aim to use this radar data to estimate the skeleton of the human with the aid of CNNs.

There are multiple approaches to represent the radar reflection data. The simplest approach is a point-cloud representation of the reflection points in 3-D XYZ space, as shown in Fig.~\ref{pcl}(a). However, such representation does not provide an indication of the size of the reflecting surface. By introducing the reflection power-levels as an additional feature $I$, based on the relationship in Eqn.~\ref{equ_rcs}, we can assign an RGB weighted pixel value to the points (Fig.~\ref{pcl}(b)), resulting in a 3-D heat-map, which may serve as an input to the CNN. Considering the maximum unambiguous depth ($X_{ua}$), azimuth ($Y_{ua}$) and elevation ($Z_{ua}$) offered by the radar with resolutions ${\Delta x, \Delta y}$ and $\Delta z$, respectively, the resulting input data dimension can be represented as:
\begin{equation}
Dimension = \frac{X_{ua}}{\Delta x}\times\frac{Y_{ua}}{\Delta y}\times\frac{Z_{ua}}{\Delta z}\times3
\end{equation}
For instance, consider a radar that can detect upto 256 reflection points in a CPI. To represent the reflection data in a 5~m~$\times$~5~m~$\times$~5~m scene, with achievable resolutions of 5~cm in all the three dimensions, the input dimensions would be  100~$\times$~100~$\times$~100 pixels, each with 3 channels (RGB) corresponding to the reflection power intensity. There are a couple of challenges with this approach. Firstly, the CNN would be size and parameter intensive as the input size is extremely big. Secondly, the input data is extremely sparse (256 points in 10$^6$ pixels) and is therefore an sub-optimal representation of the features resulting in unnecessary computation expenditure.

To overcome these challenges, here is our proposed approach. The first step is to project the reflection points onto the depth-azimuth (XY) and depth-elevation (XZ) planes, respectively. A 16~$\times$~16 RGB image would then be constructed, with each pixel corresponding to a reflection point, and the RGB channels would represent the x-coordinate, y/z coordinate (depending on the projection plane) and the normalized reflection power, $I$, respectively. The pixels corresponding to no detections would be assigned with a (0,0,0) in the RGB channels. Therefore, every CPI would yeild us with two images, each with a dimension of 16~$\times$~16~$\times$~3, as shown in Fig.~\ref{pcl}(c), thus offering a significant reduction in the input size to the CNN, which in turn significantly reduces the computational complexity of the network.

\begin{figure*}[t!]
\centering
\includegraphics[width=2.0\columnwidth]{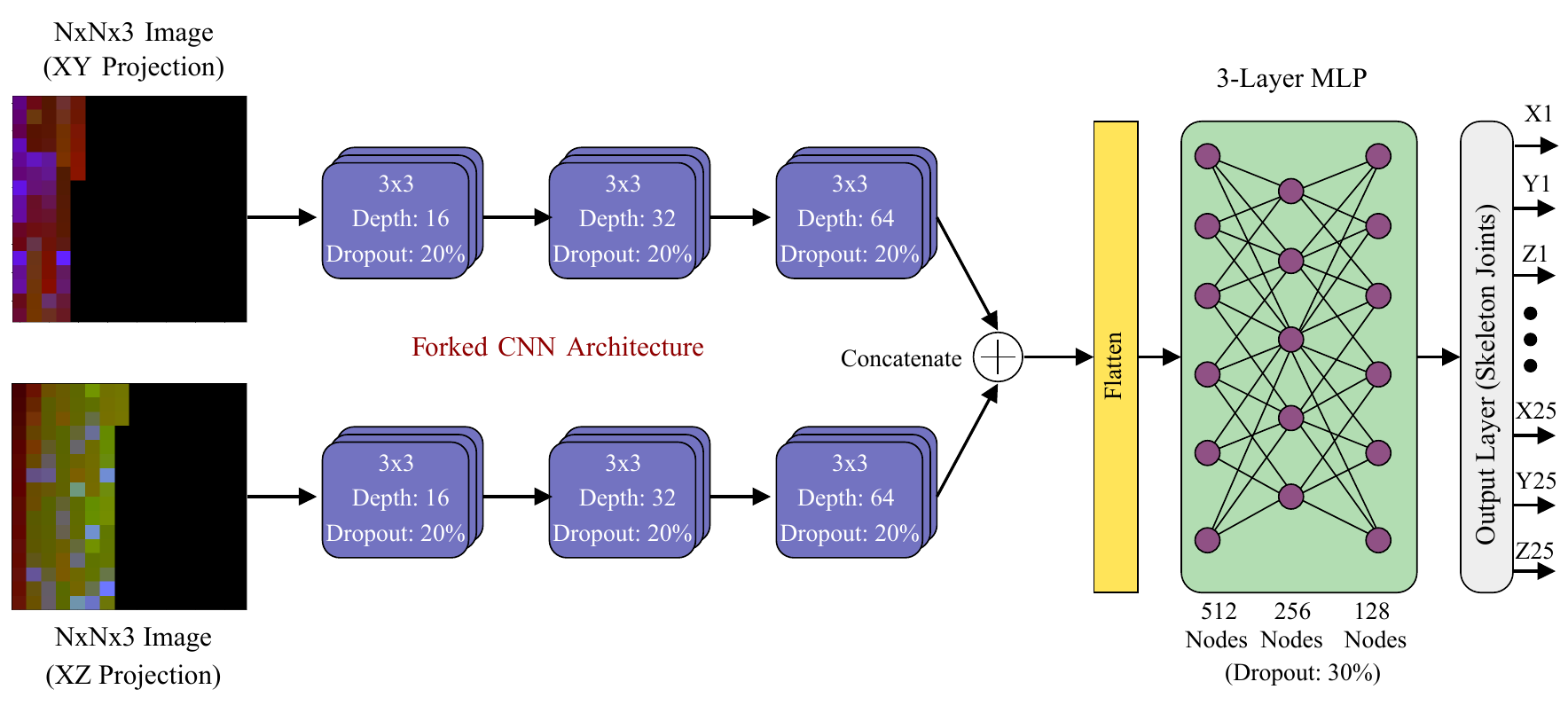}
\vspace{-0.5cm}
\caption{The N~$\times$~N~$\times$~3 image data generated from radar projections on the XY and XZ plane are subjected to a 3-layer forked CNN architecture and the outputs are concatenated and flattened. A 3-layer MLP is further used to finally obtain the X,Y and Z positions of the 25 skeletal joints from the output layer.}
\vspace{-0.5cm}
\label{cnn}
\end{figure*}

\subsection{CNN Architecture}
In this study, we use a forked CNN architecture that takes in the radar reflection data as input and provide us with the skeletal joint coordinates of the human. The advantage of using a CNN stage instead of a completely fully-connected multi-layer perceptron (MLP) is that the CNN shares weights at a given layer, thus reducing the number of trainable parameters compared to an equivalent MLP. 

Generalizing from the example scenario in the preceding section, consider the radar can detect upto $N^2$ reflection points in a CPI. After projection onto XY and YZ planes, we obtain two N~$\times$~N~$\times$~3 images with $(X,Y,I)$ and $(X,Z,I)$ as the RGB channels. If the actual number of points detected is fewer than $N^2$, the remaining pixels corresponding to no detection would be assigned with a (0,0,0) in the RGB channels. Each image is subjected to its own 3-layer CNN, with depths 16, 32 and 64 respectively. The filter size is set at 3~$\times$~3 with a single-pixel stride and same padding. The nodes are activated using a ReLU activation with a 20\% dropout to avoid overfitting. As the input dimension is small, we have not used max-pooling layers between the CNN stages, thus preserving full resolution of our data with no down-sampling operation.

The outputs from both the final CNN layers, with output dimensions of N~$\times$~N~$\times$~64 are then concatenated to form a N~$\times$~N~$\times$~128 tensor, which is then flattened and subjected to a 3-layer MLP for further flexibility in the non-linear modeling of the input (radar) - output (skeleton) relationship. The layers have 512, 256 and 128 nodes respectively, with a 30\% dropout and ReLU activation function.

In this study, we aim to map the radar reflection points to 25 distinct skeletal joints of the human body in 3-D space. Therefore, the output layer consists of 75 nodes corresponding to the $(X,Y,Z)$ locations of the 25 joints. The output layer has a linear activation function and is fully-connected to the final layer of the MLP network. The model is trained with the objective to minimize the mean-squared-error (MSE) of the predicted location of the joints with the measured ground truth. The model is trained using gradient descent using the Adam optimizer, that uses a variable learning rate depending on the rate of change of the gradient over iterations. The complete machine learning architecture is shown in Fig~\ref{cnn}.

The added advantage the proposed method offers is that this approach would not only work with radar systems that have both azimuth and elevation antenna channels, but can also be extended to radar modules that only have antenna elements in one axis (azimuth, say). In the latter case, two radars may then be used, with one capturing $XY$ data and the other rotated at 90$^o$ to capture $XZ$ data. This way N~$\times$~N~$\times$~3 images can be directly generated with no projection operation required as each radar detects the reflected points in the respective single plane. The proposed approach also eliminates the need for data association or complex construction of 4D CNNs. Finally, by incorporating the reflection power levels, we provide the CNN with an additional feature to aid the learning process and distinguish between the reflections from a larger RCS of the body (torso, say) from a smaller RCS (elbow, say). 

\section{Experiments and Results}
\subsection{Experimental Setup and Frames Association}
In this study we used Texas Instruments AWR 1642 boost mmWave radar transceivers, that has two transmit and four receive channels on a linear axis, which in its traditional orientation would resolve the radar reflection points in range (depth) and azimuth, only. We used two of these, \textbf{R-1} and \textbf{R-2} (say), with \textbf{R-2} rotated 90$^o$ counter-clockwise with respect to \textbf{R-1}, with the azimuth now corresponding to the elevation of the reflection points. Both the radars transmitted a 3.072 GHz wide chirp, centered at 79 GHz, every 92 $\mu$s. A dual-slot 3-D frame was developed and used to mount both the radars to ensure stability and consistent data collection. The processed radar point cloud data from both radars was captured via USB cables on a robot operating system (ROS) interface, running on a Linux computer. Each radar would return upto 256 detected points in a coherent processing interval, including their position (depth, elevation/azimuth), velocity and intensity at 20 frames-per-second(fps). Every return also carried a header with the UTC time-stamp and the radar module index.

To capture the ground truth data, we used a Microsoft Kinect connected to a Windows computer, using a MATLAB API. The infra-red (IR) sensor data coupled with the Mathworks developed skeletal tracking algorithm provided us with the depth, azimuth and elevation information of 25 joint positions, as well the UTC time-stamp in each frame. A common time server was used to synchronize clocks on both the computers capturing data (Radar and Kinect), with the clock slew in the order of one millisecond, which was tolerable. The UTC time-stamps from Kinect and radar frames were used for frame identification and association.

The experiment was setup in an open space in the Electrical and Computer Engineering department at the University of Arizona. Two human subjects with varying sizes were used, one-at-a-time, to collect the data. The subjects performed four different actions in contiguous sets, viz. (i) Walking, (ii) Left-Arm Swing, (iii) Right-Arm Swing, and (iv) Both-Arms Swing. We acquired $\approx$32000 samples of training data and $\approx$6000 samples of validation/development data set, to be used for training the model. $\approx$1700 samples of test data was also collected with the human subject perform the four actions, in no ordered fashion, for added robustness.

The acquired data from both the radars was first separated using the module information from the frame headers and then associated frame-by-frame with the corresponding Kinect return using the UTC time-stamps. The radar returns in each frame were normalized in range, azimuth (\textbf{R-1}), elevation (\textbf{R-2}) and intensity corresponding to the dimensions of the experiment space. The normalized data was then used to generate two RGB images every frame, corresponding to \textbf{R-1} and \textbf{R-2} respectively, based on the approach described in Section IV. Note that we did not have the plane projection stage as \textbf{R-1} and \textbf{R-2} already provided returns in XY and XZ planes respectively. The ground truth skeletal joint positions obtained using Kinect were also normalized to a [0,1] range. The normalization parameters were stored to rescale the predictions from the model and obtain the real-world joint locations.

\begin{figure}[t!]
 \centering
\includegraphics[width=\columnwidth]{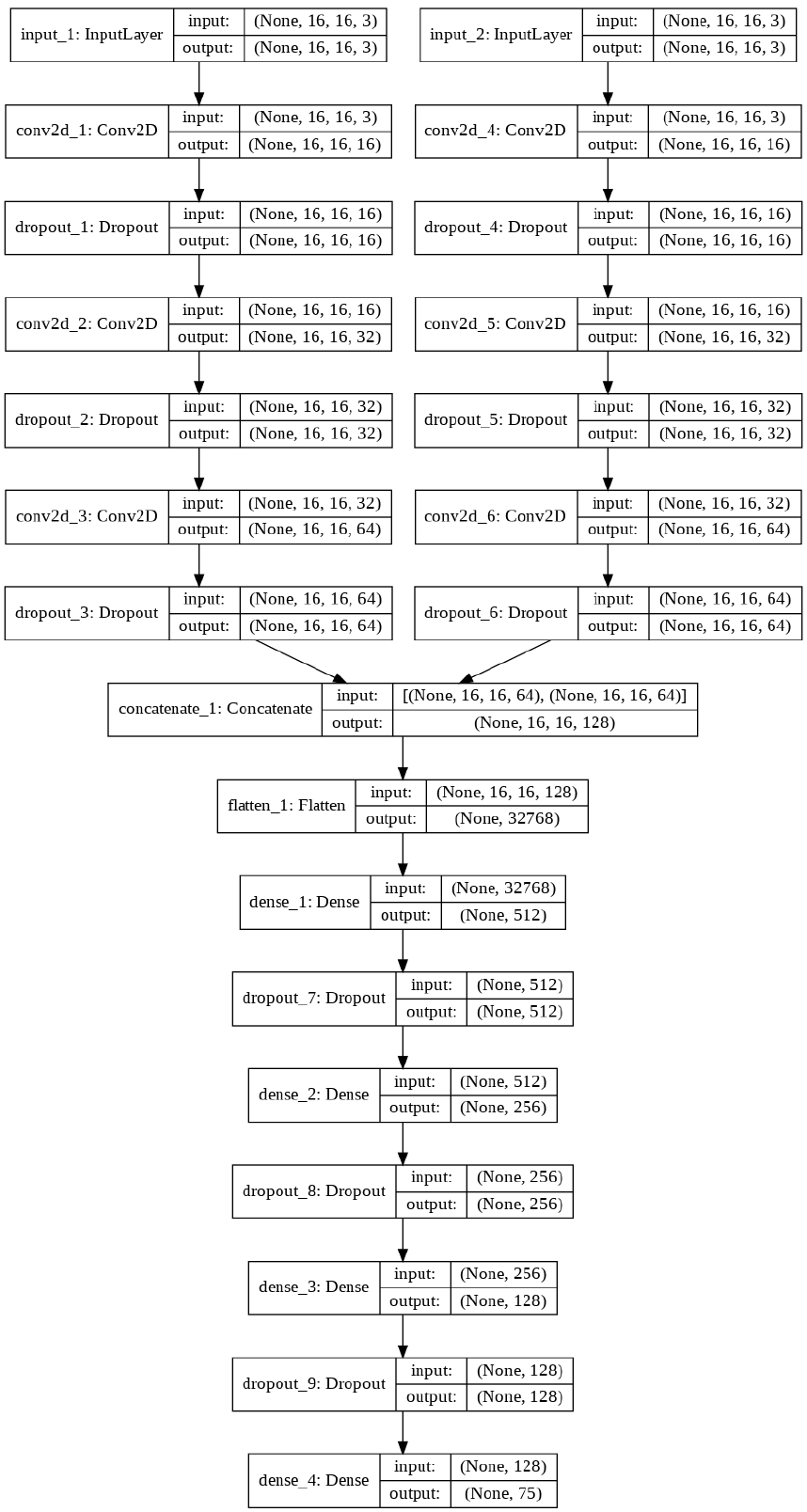}
\caption{Signal flow graph of \textit{mm-Pose} architecture built on a Keras framework }
\vspace{-0.5cm}
\label{arch}
\end{figure}

\subsection{Training the Architecture} 

A forked-CNN architecture, as described in Section IV was used as our learning algorithm. The primary reason to use CNNs in this study, as opposed to a complete multi-layer-perceptron (MLP) deep-network was to reduce the computational complexity of the network and achieve real-time implementation. Note that unlike traditional CNN layers in classification problems that are aimed to learn and generate spatial filters (edge/corner detection) in higher dimensional space, we have only used them to map our RGB encoded radar data to a higher dimensional representation, at a lower complexity, on account of CNNs' inherent property of using shared trainable weights in a kernel. Moreover, although the pixels in the encoded RGB image have relationships in terms of the range, there is no associated visual meaning by just physical inspection. Therefore, the resulting activations at intermediate layers would also have no visual interpretability and are therefore excluded for presentation in this paper.

\begin{figure}[t!]
\centering
\includegraphics{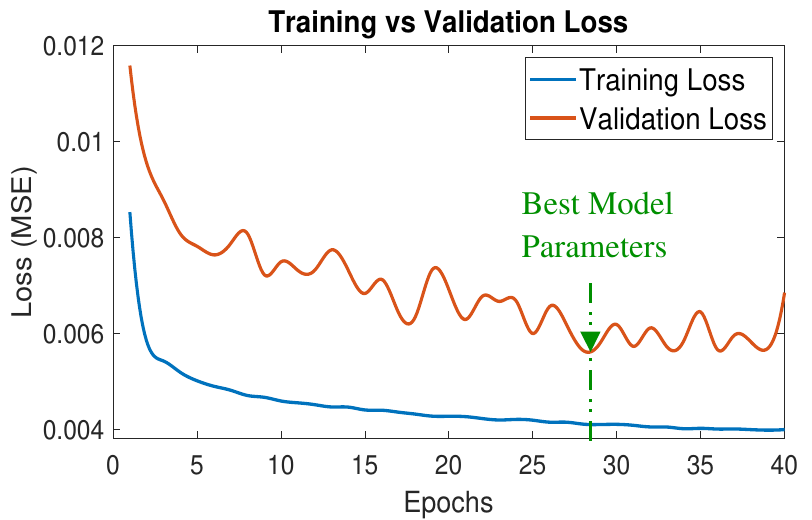}
\vspace{-0.5cm}
\caption{The training/validation loss curves for \textit{mm-Pose}. We choose the best model parameters before the model starts to overfit. }
\vspace{-0.5cm}
\label{loss}
\end{figure}

\begin{figure}[b!]
\centering
\includegraphics{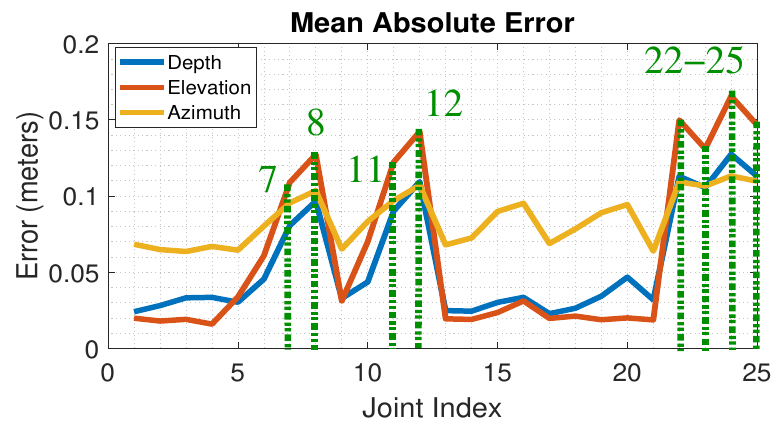}
\vspace{-0.25cm}
\caption{The MAE of \textit{mm-Pose} predicted 25 joint locations (in all three dimensions) over all the frames in the test data set. Note that the 6 outlier joint indices that offer the highest MAEs have been highlighted in green. }
\label{mae}
\end{figure}

\begin{figure*}[t!]
\centering
\includegraphics[width=2.0\columnwidth]{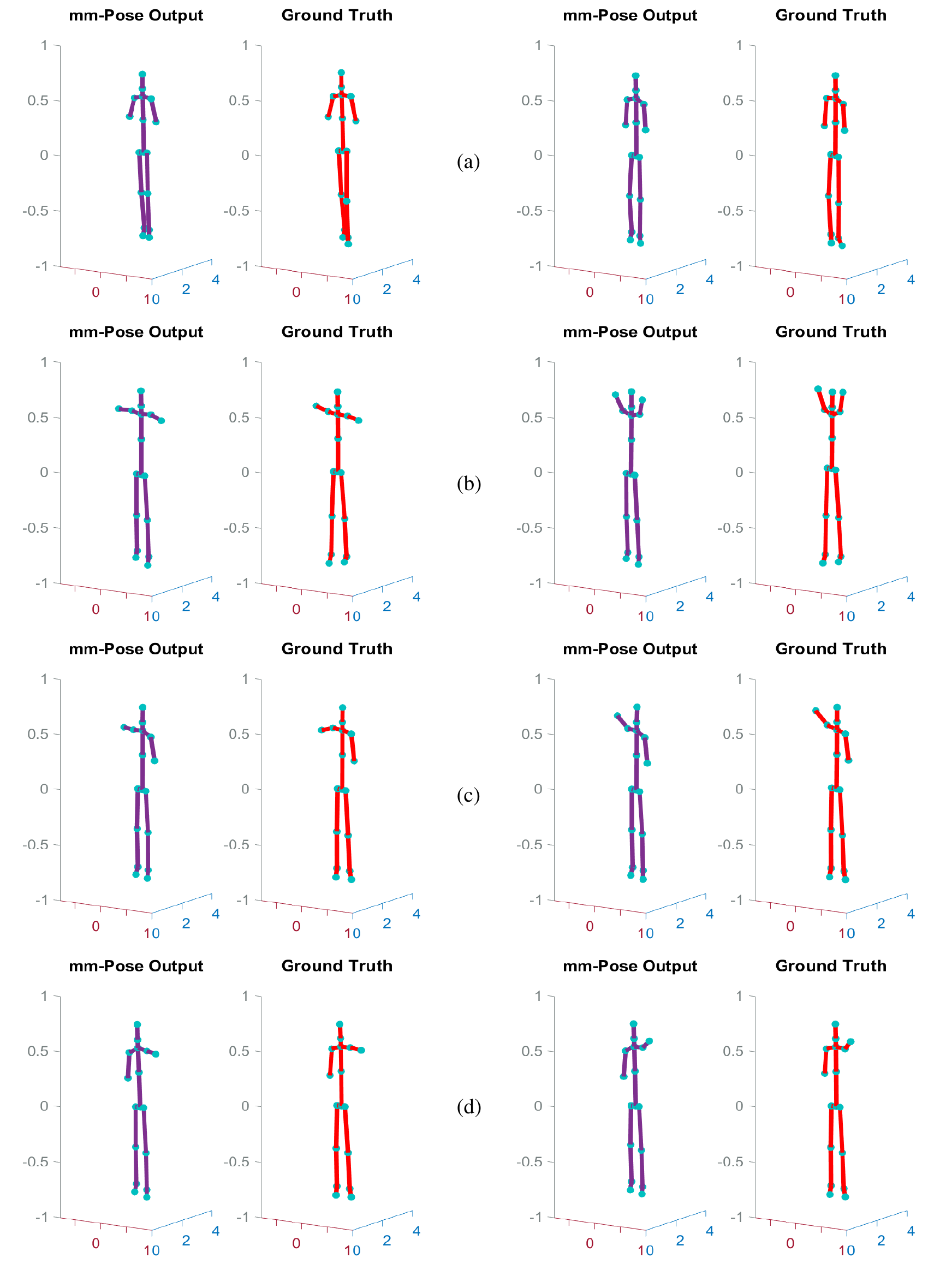}
\vspace{-0.25cm}
\caption{Visual representation of the 17-points \textit{mm-Pose} vs ground truth on the testing data with two frames shown for (a) Walking, (b) Both-arms swing, (c) Right-arm swing and (d) Left-arm swing. The axes show elevation (gray), azimuth (red) and depth (blue) in meters.}
\label{skeleton}
\end{figure*}

\begin{figure*}[t!]
\centering
\includegraphics{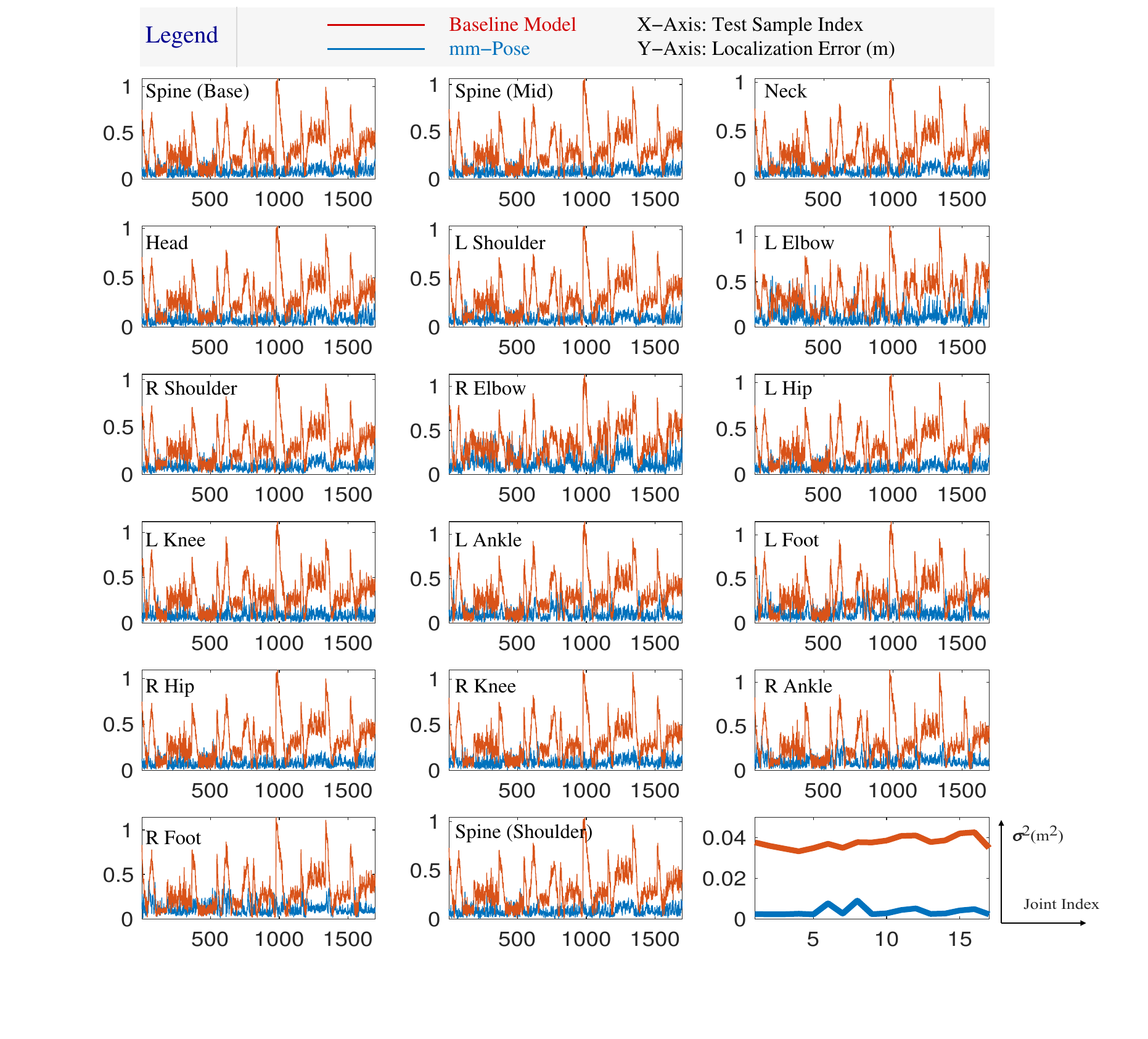}
\vspace{-1.75cm}
\caption{Comparing the localization error (in meters) of \textit{mm-Pose} (blue) vs baseline model (red) the 17 joints across 1696 frames of the testing data-set. The bottom-right figure depicts the localization error variance offered by the baseline model and the proposed \textit{mm-Pose}.}
\vspace{-0.5cm}
\label{baseline}
\end{figure*}

\begin{figure*}[t!]
\centering
\vspace{-0.5cm}
\includegraphics{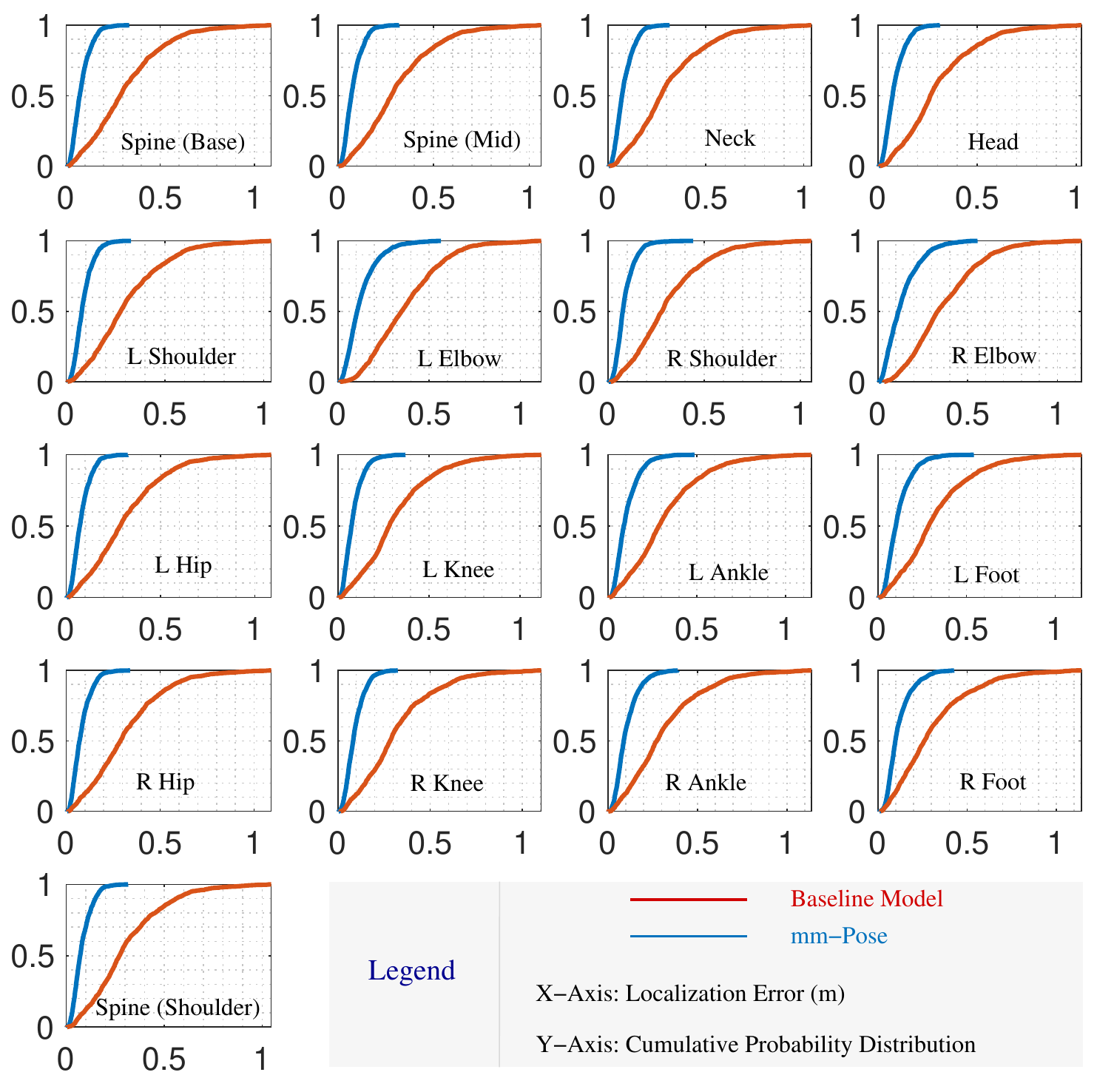}
\caption{Comparing the cumulative probability distributions of the localization errors from \textit{mm-Pose} (blue) vs baseline model (red) the 17 joints across 1696 frames of the testing data-set. \textit{mm-Pose} has a faster convergence than the baseline architecture thereby demonstrating a high probability of a lower localization error.}
\vspace{-0.5cm}
\label{baseline2}
\end{figure*}

The architecture was trained using the acquired experiment data, after normalization, on Google Colaboratory platform, powered by nVidia Tesla K-80 GPU. The model was built using Keras Functional API, built on top of Tensorflow. The signal flow graph with the input and output dimensions are shown in Fig.~\ref{arch}. The model was trained using an Adam optimizer with the objective of minimizing the mean-squared-error (MSE) of the output with respect to the ground truth data. A callback mechanism was put in place to track the training loss and the validation loss after each epoch, and save the model parameters (weights) each time the validation loss improves from the previous best logged loss. This regularization technique, coupled with dropout at each layer in the network is aimed to reduce the chances of overfitting. After training the model over several epochs, the model that yielded the best validation loss, prior to the occurrence of overfitting, is retained for further testing and evaluation, as shown in Fig.~\ref{loss}.

\subsection{Test Results and Analysis}
The trained model is evaluated in two stages. First, the testing data is used to evaluate the model's performance by evaluating the mean-absolute-error (MAE) of the predictions compared to the ground truth in depth (X), azimuth (Y) and elevation (Z) of all the 25 points. Second, the model's efficiency in terms of achieving the desired accuracy is evaluated by comparing it with a baseline architecture. As this is a relatively unexplored field of research, there is a lack of existing mmWave radar based skeletal joint databases and architectures. Therefore, to establish a baseline for comparing our results and architecture, we compare it to a model that would always produce the average location of each of the joints based on the training data.

\subsubsection{Localization Accuracy}
The MAE of all the 25 joint locations is shown in Fig~\ref{mae}. From our results, we observe that a few joint indices are outliers in the training process and offer the highest MAE. We also observed that the outliers offer a consistently high error across all the frames. The outlier joints correspond to (i) Wrist, (ii) Palm, (iii) Hand Tip and (iv) Thumb, of both left and right hands. While the ground truth data using Kinect could resolve for these joints on account of high-resolution spatial imagery, we acknowledge the challenges of representing such extremely granular and small RCS joints using mmWave radar returns alone. 

As the general skeletal representation of human pose could still be constructed with the remaining 17 points, with negligible effect on its visual interpretability, the 8 outliers, as listed above, were excluded from further analysis. The \textit{mm-Pose} predicted skeleton for two frames for each of the four postures, from  the test data, along with the ground truth for comparison, is shown in Fig~\ref{skeleton}. Also note that the outlier joint positions have also been removed from the ground truth data for consistent representation and comparison. The proposed \textit{mm-Pose} architecture offered average localization errors of 3.2 cm in depth (X), 2.7 cm in elevation (Z) and 7.5 cm in azimuth (Y), respectively. The results show that our model (17 joints) offers better localization in X and Z axis than MIT's RF-Pose3D (8 keypoints), however at a greater localization error in dimension due to a higher variance in the location of all the joints in that dimension. The localization error metrics comparing \textit{mm-Pose} with RF-Pose3D is presented in Table~\ref{locacc}.

\begin{table}[h!]
\vspace{-0.25cm}
\centering
\caption{Localization Accuracy Comparison}
\resizebox{\columnwidth}{!}{
\centering
\begin{tabular}{c|c|c|c|} 
\cline{2-4}
\multirow{2}{*}{}                   & \multicolumn{3}{c|}{Localization Accuracy}  \\ 
\cline{2-4}
                                    & Depth (X) & Elevation (Z) & Azimuth (Y)     \\ 
\hline
\multicolumn{1}{|c|}{RF-Pose3D (MIT)} & 4.2 cm    & 4.0 cm      & \cellcolor{green!25}4.9 cm            \\ 
\hline
\multicolumn{1}{|c|}{\textit{mm-Pose}}       & \cellcolor{green!25}3.2 cm    & \cellcolor{green!25}2.7 cm      & 7.5 cm            \\
\hline
\end{tabular}}
\label{locacc}
\end{table}

\subsubsection{Architecture Metrics}
The implementation goal of our proposed method is deployment for real-time application. To achieve this, smaller model size and short inference time is extremely desired. Our model makes use of CNNs, with a 3$\times$3$\times$3 kernel (27 trainable parameters) to transform the 16$\times$16$\times$3 input images (low-size with high-resolution-radar information) to every higher dimensional representation of the input data, in order to aid the learning process. The proposed model provides an average inference time of $\approx$150 $\mu s$ per frame, which make it suited for real-time implementation. A conventional multi-layer perceptron feed-forward network would require 16$\times$16$\times$3 = 768 learning units to achieve every transformative representation, with the number of multiplications between layers also increasing exponentially, and the inference time also increasing significantly as a consequence.
\par Finally, we compare \textit{mm-Pose}'s prediction and localization errors, on the testing data, with a baseline model that always outputs the mean of the training data's ground truth for each of the 17 joint locations. The 3-D euclidean distance of the predicted position and the ground truth positions are compared to evaluate our model's performance. We observe, as shown in Fig.~\ref{baseline}, that the distance between \textit{mm-Pose}'s predicted joint locations from the actual ground truth is an order of magnitude lower than the baseline model. A cumulative probability distribution of the localization error for each of the joints. as shown in Fig.~\ref{baseline2}, further elucidates \textit{mm-Pose}'s consistently lower errors with a steep convergence to the maximum, unlike the baseline architecture.   

\subsection{Practical Implementation}
The trained \textit{mm-Pose} model was then implemented for real-time human skeleton based pose estimation using mmWave radars. The entire system was achieved on a ROS interface by instantiating four sequential nodes. Node 1 was the radar node that published the point cloud information from reflection signals following the signal processing stages. Node 2 was subscribed to the published point cloud data, separated \textbf{R-1} and \textbf{R-2} frames and published the corresponding 16$\times$16$\times$3 radar data encoded RGB images. Node 3 was the mmPose node that used the RGB images from Node 2 to predict the normalized locations of all the joints. The final node (Node 4) un-normalized the predicted joint locations and converted them to a real world coordinate system before mapping it over to a display for monitoring. The system was tested out with different human subjects and the real-time pose estimation system was successfully verified.  

\subsection{Limitations}
With the unavailability of mmWave radar-skeletal databases, the data acquisition stage was the most expensive process in this study. As \textit{mm-Pose} was developed with the training data obtained while performing four different movements as listed above, the output might not be reliable if the subject performs a completely different spatial motion (crouch, bend over etc.). However, with additional training encompassing more behavioral data, future variants of mmPose could be developed to be more robust, including extension to simultaneous multiple human skeletal pose tracking.

\section{Conclusion}
In this paper, \textit{mm-Pose}, a real-time novel skeletal pose estimation using mmWave radars is proposed. The 3-D XYZ radar point cloud data (upto N$^2$ points per CPI) is first projected onto the XY and XZ planes, followed by an N$\times$N$\times$3 RGB image, with the RGB channels corresponding to the 2-D position and intensity information of each reflection point. This data representation was aimed at eliminating a voxel based learning approach and reducing the sparsity of the input data. A forked-CNN based deep learning architecture was trained to estimate the X, Y, and Z locations of 25 joints and construct a skeletal representation. 8 outlier joints were identified that did not aid to the learning process and were subsequently removed from our system and further analysis, as we were able to reasonably reconstruct the skeletal pose using the remaining 17 joints. The proposed architecture offered significant reduction in computational complexity compared to traditional MLP networks and offered a much lower localization error and variance when compared to the baseline architectures. The average localization errors of 3.2 cm in depth (X) and 2.7 cm in elevation (Z) outperforms MIT's RF-Pose3D by $\approx$24\% and $\approx$32\%, respectively. However, the localization error of 7.5 cm in azimuth (Y)  was found to be greater than the 4.9 cm offered by RF-Pose3D. The end-to-end system was verified successfully for real-time estimation, using mmWave radars and the proposed \textit{mm-Pose} architecture on ROS. The current implementation of mmPose was developed with the data obtained using four different motions, however, more motions could be added by the rather expensive process of data collection and labeling for a wide range of spatial motions for added robustness. Finally, \textit{mm-Pose} could be used for a wide range of applications including (but not limited to) pedestrian tracking, real-time patient monitoring systems and through-the-wall pose estimations for military applications. 

\bibliographystyle{ieeetr}
\bibliography{references}
\end{document}